\documentclass[pra,a4paper,floatfix,superscriptaddress]{revtex4}
\usepackage{pstricks,graphicx,amsmath,bbm,mathrsfs,amssymb,times,psfrag,pifont}
\usepackage{epsfig}
\newcommand{\ket}[1]{\vert #1 \rangle} \newcommand{\bra}[1]{\langle #1 \vert}

\newcommand{\bmsigma}{\boldsymbol \sigma}
\newcommand{\bmSigma}{\boldsymbol \Sigma}
\newcommand{\bmX}{\boldsymbol X}
\newcommand{\bmA}{\boldsymbol A} 
\newcommand{\bmB}{\boldsymbol B} \newcommand{\bmC}{\boldsymbol C}
\newcommand{\bmS}{\boldsymbol S} 

\def\mT{{m_{\rm T}}}
\def\mR{{m_{\rm R}}}

\def\NT{{M_{\rm T}}}
\def\NR{{M_{\rm R}}}

\def\etaT{{\eta_{\rm T}}}
\def\etaR{{\eta_{\rm R}}}
\def\Nth{{N_{\rm th}}}

\def\FIPS{F_{\rm IPS}}
\def\FCPS{F_{\rm CPS}}
\def\FT{F_{\rm T}}
\def\Tr{{\rm Tr}}

\begin{document}

\title{Reliable source of conditional non-Gaussian states from single-mode thermal fields}

\author{A.~Allevi}
\email{alessia.allevi@uninsubria.it} \affiliation{CNISM U.d.R. Como, I-22100 Como,
Italy.}

\author{A.~Andreoni}
\affiliation{Dipartimento di Fisica e Matematica, Universit\`a degli Studi dell'Insubria,
I-22100 Como, Italy} \affiliation{CNISM U.d.R. Como, I-22100 Como, Italy.}

\author{M.~Bondani}
\affiliation{Istituto di Fotonica e Nanotecnologie, CNR, I-22100 Como, Italy.}
\affiliation{CNISM U.d.R. Como, I-22100 Como, Italy.}

\author{M.~G.~Genoni}
\affiliation{CNISM U.d.R. Milano Universit\`a, I-20133 Milano, Italy.}
\affiliation{Dipartimento di Fisica, Universit\`a degli Studi di Milano, I-20133 Milano,
Italy.}

\author{S.~Olivares}
\email{stefano.olivares@mi.infn.it} \affiliation{CNISM U.d.R. Milano Universit\`a,
I-20133 Milano, Italy} \affiliation{Dipartimento di Fisica, Universit\`a degli Studi di
Milano, I-20133 Milano, Italy.}

\begin{abstract}
We address both theoretically and experimentally the generation of pulsed non-Gaussian
states from classical Gaussian ones by means of conditional measurements. The setup
relies on a beam splitter and a pair of linear photodetectors able to resolve up to tens
of photons in the two outputs. We show the reliability of the setup and the good
agreement with the theory for a single-mode thermal field entering the beam splitter and
present a thorough characterization of the photon statistics of the conditional states.
\end{abstract}
\maketitle

\section{Introduction}
\label{s:intro}

The subtraction of photons from an optical field is of both fundamental and practical
interest, because it is linked to the properties of the annihilation operator and plays a
leading role in quantum information protocols involving non-Gaussian states generation,
manipulation and distillation. In fact, the simplest way to generate a non-Gaussian
optical state starting from a Gaussian one consists in subtracting photons from it.
Photon subtraction can be implemented by inserting a beam splitter in the optical path of
the original state, detecting the number of photons of the reflected portion and
selecting the transmitted portion only if a certain condition on the number of detected
photons is satisfied. The challenging part of this scheme is the use of detectors able to
resolve the number of photons. As a matter of fact, if, on one hand, it is nowadays quite
easy to detect a single photon (see, e.g., Ref.~\cite{Par:JPB:09} and references
therein), on the other hand, the limited availability of photon counters that can resolve
higher numbers of photons has led to the quest for indirect ways to obtain such
information \cite{rossi:PRA:04,zam:PRL:05,bri:JMO:09}.

It is worth mentioning that the subtraction of photons allows not only the generation of
non-Gaussian states, but also the enhancement of the non-locality of bipartite states
\cite{opa:PRA:00,coc:PRA:02,oli:PRA:03,inv:PRA:05,oli:LP:06}, or the generation of highly
non-classical states \cite{wen:PRL:04,oli:JPB:05} useful for quantum information purposes
\cite{cerf:05}. Nevertheless, non-Gaussianity is a necessary ingredient for
continuous-variable entanglement distillation
\cite{eisert:PRL:02,fiura:PRL:02,giedke:PRA:02} and different protocols relying on
Gaussification of entangled non-Gaussian states
\cite{browne:PRA:03,eisert:AOP:04,hage:NJP:07} or on de-Gaussification of entangled
Gaussian states have been proposed \cite{taka:09}. In all these approaches an important
role is played by photodetectors  able to perform conditional measurements.

In this paper we report a thorough analysis of a setup based on hybrid photodetectors
allowing the discrimination of the number of detected photons up to tens
\cite{bon:ASL:09,bon:OL:09}. The aim of the paper is twofold: firstly we demonstrate the
feasibility of our setup and, secondly, we investigate its reliability by characterizing
the generated conditional states. The input Gaussian states we employ to achieve these
goals are single-mode thermal fields. Thermal states are diagonal in the photon-number
basis, thus, the knowledge of their photon statistics fully characterizes them and their
conditional non-Gaussian counterparts, which are still diagonal. Thanks to this property,
we can give a complete analytical description of the behavior of our setup, including the
actual expressions of the conditional states, and we can verify the agreement between the
theoretical expectations and the experimental results with very high accuracy and
control. This is a fundamental test in view of the application of our setup to the
generation of more sophisticated states by conditioning non-classical, multipartite and
multimode ones \cite{twbrd}.

Throughout the paper we investigate two possible scenarios. We refer to the first one as
``conclusive photon subtraction'' (CPS): a photon-number resolving detector is used to
condition the signal and to {\em conclude} which is the effective number of subtracted
photons. The second one is the ``inconclusive photon subtraction'' (IPS): an ``on/off''
Geiger-like detector, i.e., a detector only able to distinguish the presence from the
absence of photons is employed, preventing us from inferring the actual number of
subtracted photons.

The paper is structured as follows.  Section \ref{s:COND:th} addresses the generation of
conditional states by means of detectors with an effective photon-number resolving power.
We discuss the model in the presence of non-unit quantum efficiency and give some
analytical results.  In Section \ref{s:IPS:th} we briefly review the IPS process on
thermal Gaussian fields; we also investigate the main properties of the generated
conditional non-Gaussian states that will turn to be useful for the characterization of
our setup. In Section \ref{s:EXP} we report the experimental demonstration of our scheme
and thoroughly characterize the obtained conditional states. Section \ref{s:concl} closes
the paper and draws some concluding remarks.

\section{Conditional non-Gaussian states from thermal fields
via conditional measurements}
\label{s:COND:th}

\begin{figure}[tb]
\includegraphics[width=0.5\columnwidth]{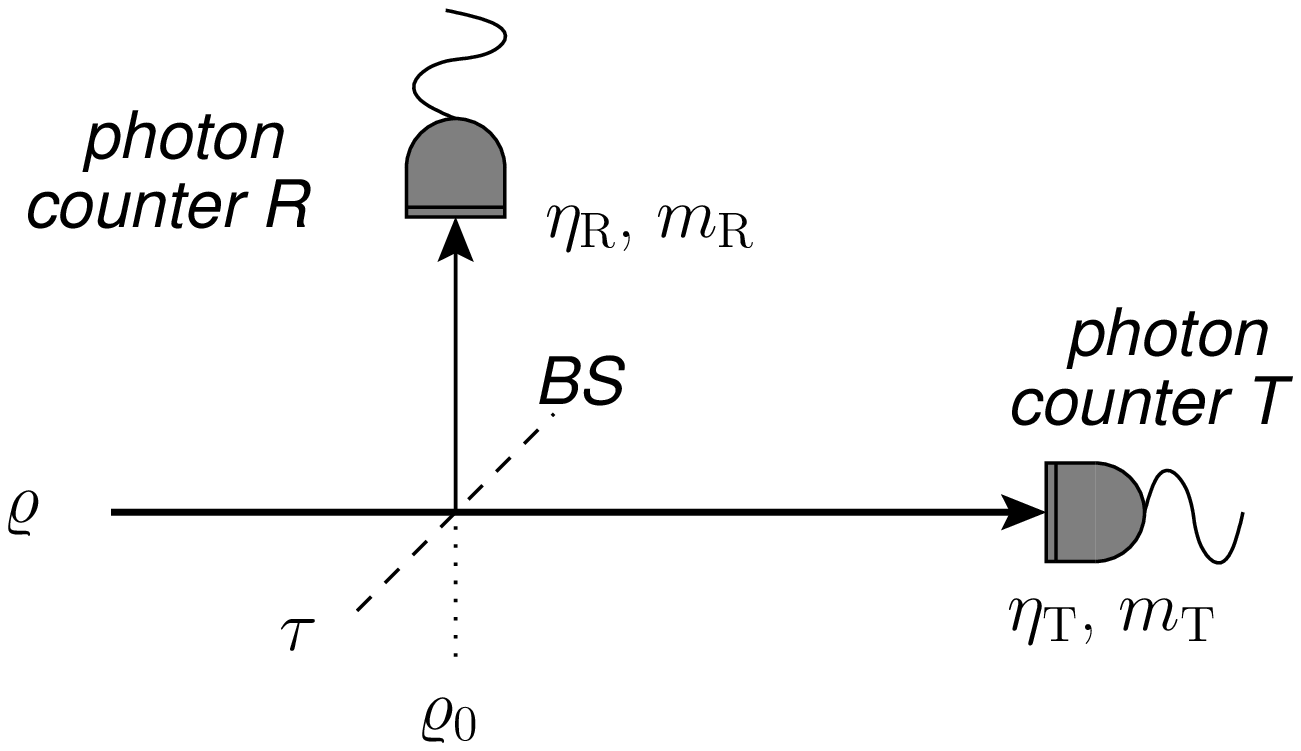}
\caption{\label{f:scheme:cond}
Scheme for the generation of conditional non-Gaussian states via
photon subtraction. A thermal input state
$\varrho$ is mixed with the vacuum state $\varrho_{0} =
\ket{0}\bra{0}$ at a beam splitter (BS) with transmissivity $\tau$.
Two photon counters (R, T) with quantum efficiency $\eta_k$,
$k={\rm R}, {\rm T}$, are used to generate and analyze conditional
states. See text for details.}
\end{figure}
In Fig.~\ref{f:scheme:cond} we depict the conditional photon-subtraction scheme based on
a beam splitter (BS) and two photon-number resolving detectors. Although in our
experimental realization we will consider only thermal states, for the sake of generality
here we consider a diagonal state of the form $\varrho = \sum_n \varrho_n
\ket{n}\bra{n}$.  After the evolution through the BS with transmissivity $\tau$, the
initial two-mode state $R_0=\varrho\otimes \ket{0}\bra{0}$ is transformed into the state
\begin{equation}\label{evolved}
R =
\sum_{n=0}^{\infty}\varrho_n \sum_{k,l=0}^n A_k^n(\tau) A_l^n(\tau)
\ket{n-k}\bra{n-l}\otimes \ket{k}\bra{l},
\end{equation}
where $A_s^n(\tau) = \sqrt{ {n \choose s} \tau^{n-s} (1-\tau)^s}$. Then, the reflected
part of the beam undergoes measurement. The positive-operator valued measure (POVM)
describing a realistic photon counting device with quantum efficiency $\eta$ is given by
\cite{DC:note}
\begin{equation}\label{POVM:real}
\Pi_m(\eta) = \sum_{s=m}^{\infty} B_{s,m}(\eta) \ket{s}\bra{s},
\end{equation}
in which $B_{s,m}(\eta) =  {s \choose m}  \eta^m (1-\eta)^{s-m}$. If the photon-counter
in the reflected beam detects $\mR$ photons, the conditional photon subtracted (CPS)
state obtained in the transmitted beam is:
\begin{align}\label{rho:CPS}
\varrho_{\rm CPS} (\mR)&= \frac{1}{p_{\rm R}(\mR)}
{\rm Tr}_{\rm R}[R\, {\mathbbm I}\otimes\Pi_\mR(\etaR) ]\nonumber\\
&= \frac{1}{p_{\rm R}(\mR)} \sum_{s=\mR}^{\infty} \sum_{n=0}^{\infty} B_{s,\mR}(\etaR)\nonumber\\
&\hspace{1.5cm}\times
\varrho_{s+n} \left[ A_s^{s+n}(\tau) \right]^2 \ket{n}\bra{n},
\end{align}
where $\eta_k$ is the quantum efficiency of the detector located in the reflected
($k={\rm R}$) and in the transmitted ($k={\rm T}$) beam paths, respectively. Note that
the state in Eq.~(\ref{rho:CPS}) is still diagonal. The overall probability $p_{\rm
R}(\mR)$ of measuring $\mR$ in the reflected beam reads:
\begin{equation}
p_{\rm R}(\mR) =  \sum_{s=\mR}^{\infty} \sum_{n=0}^{\infty} B_{s,\mR}(\etaR)
\varrho_{n+s} \left[ A_s^{n+s}(\tau) \right]^2,
\end{equation}
which represents the  marginal distribution of the joint probability,
\begin{align}
p_{\rm TR}(\mT,\mR) &= {\rm Tr}[ R\, \Pi_\mT(\etaT) \otimes \Pi_\mR(\etaR)]\nonumber\\
&= \sum_{t=\mT}^{\infty} \sum_{s=\mR}^{\infty}
B_{s,\mR}(\etaR)\,B_{t,\mT}(\etaT) \nonumber \\
&\hspace{1.0cm}\times \left[ A_s^{s+t}(\tau) \right]^2 \, \varrho_{s+t} ,
\label{pj:no:dc}
\end{align}
that detectors T and R measure $\mT$ and $\mR$ photons, respectively. By taking $\varrho$
in a single-mode thermal state $\nu(\Nth)$,
\begin{align}
\nu(\Nth) &= \sum_{n=0}^{\infty}
\nu_n(\Nth) \ket{n}\bra{n}, \label{th:st}\\
\nu_n(\Nth) &= \frac{1}{1+\Nth}
\left(\frac{\Nth}{1+\Nth}\right)^n,\label{th:st:n}
\end{align}
where $\Nth$ denotes the mean number of thermal photons, Eq.~(\ref{pj:no:dc}) reduces to
\begin{align}\label{eq:joint}
p_{\rm TR}(\mT &, \mR) = {\mT+\mR \choose \mR}
\frac{\NT^{\mT}\, \NR^{\mR}}{(1+\NT+\NR)^{\mT+\mR+1}},
\end{align}
$\NT = \tau\, \etaT\, \Nth$ and $\NR = (1-\tau)\, \etaR\, \Nth$ being the mean numbers of
detected photons of the transmitted and reflected beams, respectively.

Given $\mR$ and $p_{\rm R}(\mR)$, the conditional state $\varrho_{\rm CPS}$ in
Eq.~(\ref{rho:CPS}) can be obtained straightforwardly. From Eq.~(\ref{rho:CPS}) we can
then evaluate the Fano factor
\begin{align}
\FCPS &= \frac{ \sigma^2(M_{\rm CPS})}
{M_{\rm CPS}},
\end{align}
which is the ratio between the variance $\sigma^2(M_{\rm CPS})$ and the mean number
$M_{\rm CPS}$ of the photons detected in the CPS state. As we will see below, $\FT \ge
\FCPS \ge 1$, where $\FT = 1+\NT$ is the Fano factor of the single-mode thermal field of
the (unconditional) transmitted beam. Note that $\varrho_{\rm CPS}$ is always
super-Poissonian, which is consistent with the classical nature of the field.

To deeply characterize the output conditional state we
evaluate its non-Gaussianity. Since the state has the form
$\varrho_{\rm CPS}=\sum_n p_n |n\rangle\langle n|$, the non-Gaussianity (nonG)
measure \cite{non:g} can be written as:
\begin{equation}
\delta[\varrho_{\rm CPS}] = S[\nu(N_{\rm CPS})] + \sum_n p_n \log p_n
\end{equation}
where $N_{\rm CPS}$ is the mean photon number of $\varrho_{\rm CPS}$, and $S[\nu(N)]= N
\log(1+1/N) + \log(1+N)$ is the entropy of the thermal state $\nu(N)$.

However, due to the inefficient detection, we cannot
reconstruct the actual photon number distribution $p_n$, but only the
\emph{detected} photon number distribution $q_\mT=p_{\rm TR}(\mT,\: \mR)$
given in Eq. (\ref{eq:joint}),  where $\mR$ is the conditioning value,
and $\mT$ is the number of detected photons. Thus we can evaluate
the quantity
\begin{equation}
\varepsilon[\varrho_{\rm CPS}]= S[\nu(M_{\rm CPS})]
+ \sum_\mT q_\mT \log q_\mT \leq \delta[\varrho_{\rm CPS}] .
\end{equation}
The last inequality follows from the fact that the inefficient detection may be described
by a Gaussian lossy channel that does not increase the non-Gaussianity, followed by an
ideal (i.e., unit quantum  efficiency) detection (see Appendix \ref{appA} for details).
The quantity $\varepsilon[\varrho_{\rm CPS}]$, which can be easily evaluated from our
experimental data, turns out to be a lower bound for the actual non-Gaussianity, that is,
significant values of $\varepsilon[\varrho_{\rm CPS}]$ correspond to more markedly
non-Gaussian states $\varrho_{\rm CPS}$.

\section{Inconclusive photon subtraction on thermal states}
\label{s:IPS:th}

The conditional states introduced in the previous Section can be generated only if the
detector in the reflected beam path is able to resolve the number of incoming photons. In
this Section we consider a scenario in which the detector R (see
Fig.~\ref{f:scheme:cond}) can only distinguish the presence from the absence of light
(Geiger-like detector): we will refer to this measurement as inconclusive, as it does not
resolve the number of the detected photons. When the detector clicks, an unknown number
of photons is subtracted from $\varrho$ and we obtain the IPS state $\varrho_{\rm IPS}$.
To characterize this class of conditional state, we use the phase-space description of
the system evolution, that allows a simpler analysis with respect to that based on the
photon number basis.

The phase-space description of the IPS operated on single-mode Gaussian states can be
obtained by generalizing the analysis given in Ref.~\cite{oli:JPB:05}. The Wigner
function of the thermal state in Eq.~(\ref{th:st}) reads as follows (in Cartesian
notation):
\begin{align}
W_{\rm th}(\bmX) =
\frac{\exp\left(-\hbox{$\frac12$} \bmX^T \bmsigma_{\rm th}^{-1}\bmX \right)}
{2\pi \sqrt{\hbox{Det}[\bmsigma_{\rm th}]}},
\end{align}
where:
\begin{equation}\label{CM:thermal}
\bmsigma_{\rm th} \equiv \bmsigma_{\rm th}(\Nth) =
\frac{1+2\Nth}{2}\,{\mathbbm 1}
\end{equation}
is the covariance matrix (CM), ${\mathbbm 1}$ being the $2\times 2$ identity matrix.
According to \cite{oli:JPB:05}, the action of the BS transforms the CM of the two-mode
input state (thermal$+$vacuum)
\begin{equation}
\bmsigma_{\rm in} = \left(
\begin{array}{c|c}
\bmsigma_{\rm th} & {\boldsymbol 0} \\
\hline
{\boldsymbol 0} & \bmsigma_0
\end{array}
\right)\, ,
\end{equation}
as follows \cite{FOP:napoli:05}:
\begin{equation}
\bmsigma' \equiv
{\bmS}_{\rm BS}^T(\tau)\, \bmsigma_{\rm in}\,{\bmS}_{\rm BS}(\tau) \equiv
\left(\
\begin{array}{c|c}
\bmA & \bmC\\
\hline
\bmC^T & \bmB
\end{array}
\right)\,,
\end{equation}
where $\bmA$, $\bmB$, and $\bmC$ are $2 \times 2$ matrices and
\begin{equation}
\bmS_{\rm BS}(\tau) =
\left(\
\begin{array}{c|c}
\sqrt{\tau}\, \mathbbm{1} & \sqrt{1-\tau}\, \mathbbm{1} \\
\hline
-\sqrt{1-\tau}\, \mathbbm{1}  & \sqrt{\tau}\, \mathbbm{1}
\end{array}
\right)\,
\end{equation}
is the symplectic transformation associated with the evolution operator
$U_{\rm BS}$ of the BS.

The probability $p_{\rm on} = p_{\rm on}(r,\tau,\etaR)$ that the
on/off detector endowed with quantum efficiency $\etaR$ clicks
is given by \cite{FOP:napoli:05}:
\begin{align}
p_{\rm on} &= 1 - p_{\rm off}(r,\tau,\etaR) \\
&=1 - \left(\etaR\sqrt{{\rm Det}[\bmB + \bmsigma_{\rm M}]}\right)^{-1} \\
&= \frac{\etaR (1-\tau) \Nth}{1 + \etaR (1-\tau) \Nth}\,,
\label{click:prob}
\end{align}
where $ p_{\rm off}$ is the probability of a non-click event and
\begin{equation}
\bmsigma_{\rm M} = \frac{2 - \etaR}{2\etaR} \mathbbm{1}.
\end{equation}
The Wigner function associated with the IPS state $\varrho_{\rm IPS}$ reads:
\begin{align}\label{IPS:th}
W_{\rm IPS}(\bmX) = \frac{W_a(\bmX) - p_{\rm off}\,W_b(\bmX)}{p_{\rm on}},
\end{align}
where
\begin{equation}\label{w:f:k}
W_k(\bmX) = \frac{\exp\left(-\hbox{$\frac12$}
\bmX^T \bmSigma_{k}^{-1}\bmX \right)}
{2\pi \sqrt{\hbox{Det}[\bmSigma_k]}},\quad (k=a,b)
\end{equation}
$\bmSigma_a = \bmA$ and $\bmSigma_b = \bmA - \bmC (\bmB -\bmsigma_{\rm M})\bmC^T$. Note
that the IPS, being it the linear combination of two Gaussian functions, is no longer
Gaussian: for this reason the IPS process is also referred to as a {\em
de-Gaussification} process \cite{wen:PRL:04}. The Wigner functions in Eq.~(\ref{w:f:k})
are those of two thermal states $\nu(N_k)$ with mean number of photons $N_k$ given by
\begin{equation}\label{N:a:b}
N_a = \tau \Nth , \quad
N_b = \frac{\tau \Nth}{1 + \etaR (1 - \tau) \Nth},
\end{equation}
respectively; thus, the density matrix associated with (\ref{IPS:th}) can be written as:
\begin{equation}\label{IPS:th:dm}
\varrho_{\rm IPS} = \frac{\nu(N_a) - p_{\rm off}\,\nu(N_b)}{p_{\rm on}},
\end{equation}
and the corresponding conditional distribution of the detected photons is:
\begin{equation}\label{IPS:phs}
p_T(\mT) = \frac{\nu_{\mT}( M_a) - p_{\rm off}\,\nu_{\mT}( M_b)}{p_{\rm on}},
\end{equation}
where $M_a=\etaT N_a$ and $M_b=\etaT N_b$, being $\etaT$ the quantum efficiency of the
photon-resolving detector of the IPS state, and the $\nu(N_k)$ are given by
Eq.~(\ref{th:st:n}).

Starting from the above results, we can give some further detail about the IPS thermal
state in Eq.~(\ref{IPS:th:dm}).  The mean number of detected photons is
\begin{align}\label{N:IPS}
M_{\rm IPS} =  \frac{M_a - p_{\rm off}\,M_b}{p_{\rm on}},
\end{align}
and the variance $\sigma^2(M_{\rm IPS})$ is:
\begin{align}
\sigma^2(M_{\rm IPS}) =&\,  \frac{M_a(1+M_a) -
p_{\rm off}\,M_b(1+M_b)}{p_{\rm on}}
\nonumber\\ & - \frac{p_{\rm off}(M_a-M_b)^2}{p_{\rm on}^2}.
\end{align}
Moreover, as $M_a \ge M_b$, the Fano factor $\FIPS$ is:
\begin{align}
\FIPS &= \frac{\sigma^2(M_{\rm IPS})}
{M_{\rm IPS}} \\
&= 1 + M_b + 2\frac{M_a(M_a - M_b)}{M_a - p_{\rm off}\, M_b}
- \frac{M_a - M_b}{1 - p_{\rm off}} \ge 1,
\end{align}
in which the final inequality can be checked by substituting the actual expressions of
$M_a$, $M_b$ and $p_{\rm off}$. The state is always super-Poissonian (also in this case,
as one would expect, $\FT \ge \FIPS \ge 1$). As for the conditional states $\varrho_{\rm
CPS}$, also in this case we can characterize the non-Gaussianity of the state
$\varrho_{\rm IPS}$ from the experimental data by evaluating the quantity
\begin{equation}
\varepsilon[\varrho_{\rm IPS}]= S(\nu(M_{\rm IPS})) + \sum_\mT q_\mT \log q_\mT\; ,
\end{equation}
which is still a lower bound for the non-Gaussianity measure, \emph{i.e.}
$\varepsilon[\varrho_{\rm IPS}] \leq \delta[\varrho_{\rm IPS}]$, as explained in the
previous Section and in the Appendix \ref{appA} in more detail.

\section{Reliable source of non-Gaussian states}\label{s:EXP}
\subsection{Experimental setup}\label{s:setup}
\begin{figure}[tb]
\includegraphics[scale=0.5,angle=270]{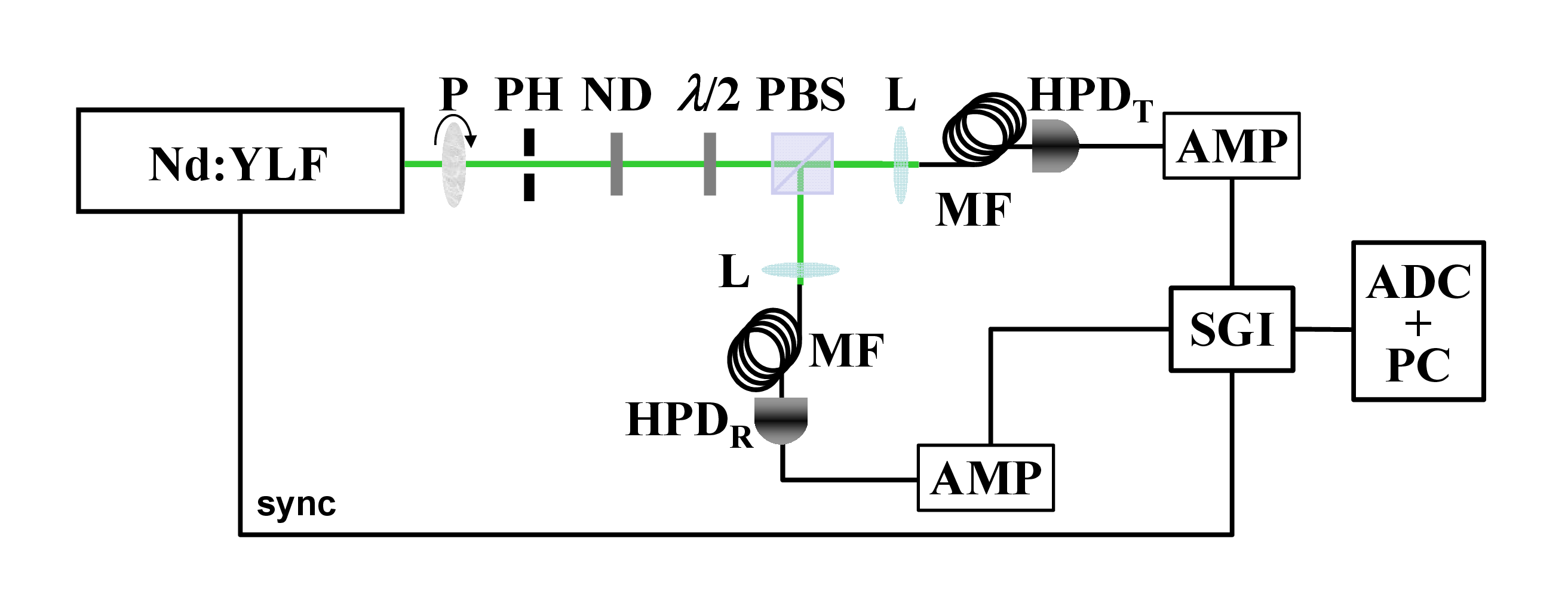}
\caption{\label{setup} (Color online) Scheme of the experimental
setup: P, rotating ground glass plate; PH, pin-hole; ND, continuously
variable neutral density filter;  $\lambda$/2, half-wave plate; PBS,
polarizing cube beam-splitter; L, collective lens; MF, multi-mode
fiber; HPD$_{\rm R,T}$, hybrid photodetector;  AMP, preamplifier plus
amplifier; SGI, synchronous-gated integrator; ADC+PC,
analog-to-digital converter.}
\end{figure}

We produced a single-mode pseudo-thermal state by inserting a rotating ground glass plate
in the pathway of a coherent field, followed by a pin-hole to select a single coherence
area in the far-field speckle pattern (see Fig.~\ref{setup}). As most detectors have the
maximum quantum efficiency in the visible spectral range, we chose to exploit the
second-harmonic linearly polarized pulses ($\lambda=523$~nm, 5.4~ps pulse duration) of a
Nd:YLF mode-locked laser amplified at 500 Hz. The thermal light was split into two parts
by a polarizing cube beam-splitter (PBS) whose transmissivity $\tau$ can be continuously
varied by means of a half-wave plate ($\lambda$/2 in Fig.~\ref{setup}). We balanced the
two exiting arms of the PBS to achieve $\tau\simeq0.5$.
The light exiting the PBS was focused in two multi-mode fibers and delivered to two
hybrid photodetectors (HPD$_{\rm R,T}$, mod. R10467U-40 Hamamatsu), endowed not only with
a partial photon-resolving capability, but also with a linear response up to 100 incident
photons. The outputs of the detectors were amplified (preamplifier A250 plus amplifier
A275, Amptek), synchronously integrated (SGI, SR250, Stanford), digitized (ATMIO-16E-1,
National Instruments) and processed offline. To analyze the outputs we model the
detection process as a Bernoullian convolution and the overall amplification/conversion
process through a very precise constant factor $\gamma$, which allows the shot-by-shot
detector output to be converted into a number of detected photons \cite{JMO09}. The
calibration procedure required performing a set of measurements of the light at different
values of the overall detection efficiency of the apparatus, $\eta$, set by rotating a
continuously variable neutral density filter wheel placed in front of the $\lambda$/2
plate. For each value of $\eta$, we recorded the data from $30\, 000$ subsequent laser
shots. For the results presented in the following we obtained the values $\gamma_{\rm
R}=0.104$~V and $\gamma_{\rm T}=0.093$~V for the calibration of the detection chains in
the reflected and transmitted arms of the beam splitter, respectively. These values of
$\gamma_{\rm R,T}$ were used to convert the voltages into number of detected photons that
were finally re-binned into unitary bins to obtain probability distributions. Once
checked the reliability of the calibrations from the quality of these distributions, the
voltage outputs of the HPD$_{\rm R,T}$ detectors were associated with numbers $m_{\rm R}$
and $m_{\rm T}$ in real time. The linearity of the detectors and the absence of
significant dark counts make our system suitable for making experiments in both the CPS
and IPS scenarios to produce conditional states. In the case of IPS, we only distinguish
the HPD$_{\rm R}$ outputs giving $m_{\rm R}=0$ from those giving any $m_{\rm R}\geq 1$ to
mimic the behavior of a Geiger-like detector.

\subsection{Conditional non-Gaussian states}\label{s:CPS:exp}

The good photon-resolving capability of HPD detectors and their linearity make it
possible to implement the CPS scheme described in Section \ref{s:COND:th}. Conditional
measurements in the reflected beam irreversibly modify the states measured in the
transmitted arm and in particular make them non-Gaussian.

To better understand the power and the limits of this kind of conditioning operation, we
follow two different approaches: first of all we fix the energy $\Nth$ of the initial
thermal field and characterize the CPS state as a function of the conditioning value
$\mR$. Secondly, we consider the properties of CPS states as a function of the mean
incoming photons $\Nth$ for a particular choice of the number $\mR$ of photons detected
in the reflected arm.  The final aim is the production of non-Gaussian states with well
defined conditioning value.

We start by presenting the results obtained by choosing a set of measurements having
$M_{\rm T} \approx 1.254$.

\begin{figure}[tb]
\includegraphics[width=0.5\columnwidth]{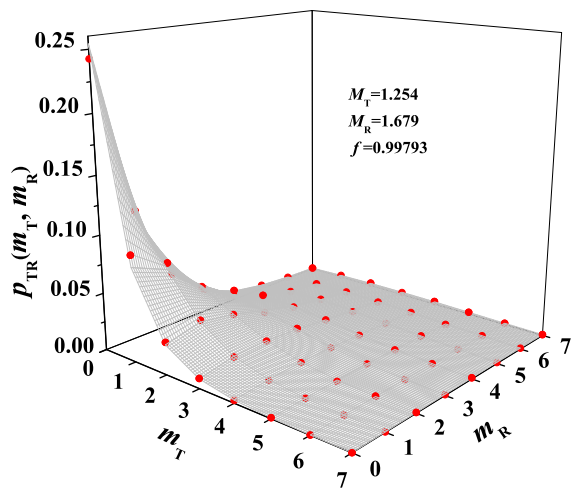}
\caption{\label{Fig:pjoint} (Color online) Joint probability
$p_{\rm TR}(\mT,\mR)$ to measure $\mR$ photons in the reflected beam and
$\mT$ photons in the transmitted one. The experimental data (red dots)
are plotted together with the theoretical surface (gray mesh). The other
involved experimental values are $M_{\rm R}=1.679$ and $M_{\rm T}=1.254$.}
\end{figure}
The joint probability $p_{\rm TR}(\mT,\mR)$ of measuring $\mR$ photons in the reflected
arm and $\mT$ photons in the transmitted one is plotted in Fig.~\ref{Fig:pjoint} as dots
together with the theoretical surface to which they perfectly superimpose.  Of course,
starting from the theoretical joint probability, we can calculate the expected
photon-number distribution of the states obtained by performing different conditional
measurements in the reflected arm [see Eq.~(\ref{rho:CPS})] and thus evaluate all the
quantities necessary to characterize the CPS states.

\begin{figure}[tb]
\includegraphics[width=0.5\columnwidth]{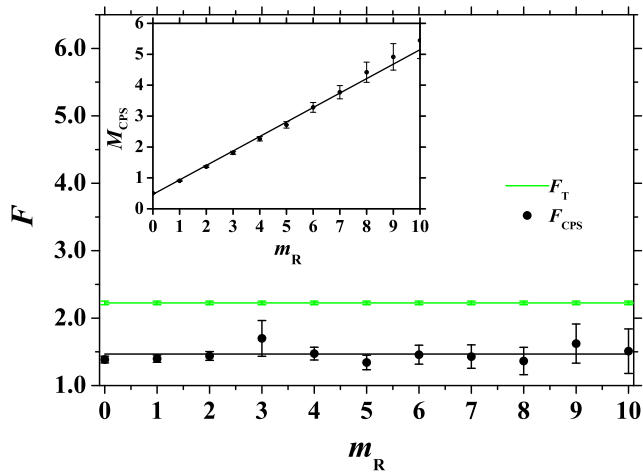}
\caption{\label{Fig:Fano27} (Color online)
Fano factors $\FCPS$ of the conditional states: experiment (black dots)
and theory (solid black line) as functions of the conditioning value $\mR$.
The green line corresponds to the Fano factor $\FT = 1+\NT$ of the (unconditional)
transmitted state. The inset refers to mean number of detected photons $M_{\rm CPS}$
of the CPS states as a function of the conditioning value $\mR$: experimental data
(dots) and theoretical curve (solid line). The values of the other involved
parameters are $M_{\rm R}=1.679$ and $M_{\rm T}=1.254$.}
\end{figure}
In Fig.~\ref{Fig:Fano27} we plot the behavior of the mean number of photons $M_{\rm CPS}$
of the conditional states and their Fano factors $\FCPS $ as a function of the different
conditioning values $\mR$. We find that the Fano factor does not depend on the particular
choice of the conditioning value $\mR$, in agreement with the analytical result
calculated from Eq.~(\ref{eq:joint}):
\begin{equation}
\FCPS = \frac{1+M_{\rm T}+M_{\rm R}}{1+M_{\rm R}} \approx 1.468.
\end{equation}
Note that the obtained value is definitely lower than that of the unconditional state,
$\FT \approx 2.225$.

The photon-number distributions of the conditional states look quite different from each
other.
\begin{figure}[tb]
\includegraphics[width=0.5\columnwidth]{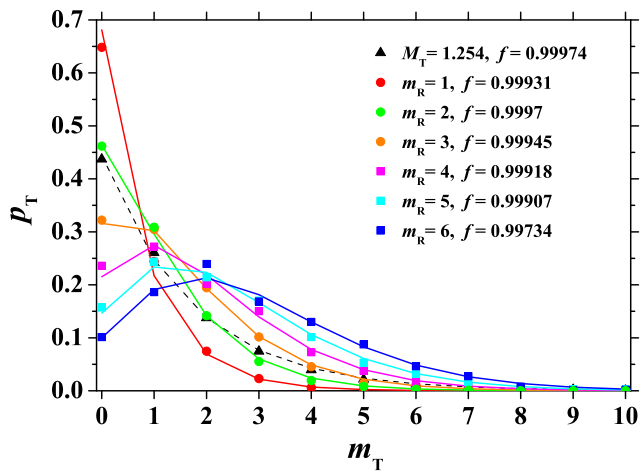}
\caption{\label{Fig:statistics27} (Color online) Reconstructed photon-number
distributions of the (unconditional) thermal state with mean number of photons
$M_{\rm T}=1.254$ (black triangles) and of the conditional states for six
different conditioning values $\mR$ (colored dots and squares). The theoretical
curves are plotted as lines according to the same choice of colors. The
corresponding fidelity $f$ is also reported.}
\end{figure}
As it is shown in Fig.~\ref{Fig:statistics27}, the larger the conditioning value, the
more different the statistics of the conditional state (colored symbols + lines) is from
that of the incoming one (black triangles + dashed line).  We note that, due to the
limited number of recorded shots (only $30\, 000$ laser-shots), the experimental points
tend to deviate from the expected $p_\mathrm{T}$ distributions at increasing conditioning
values. This behavior can be quantified by calculating the fidelity (see $f$ values
reported in Fig.~\ref{Fig:statistics27}): $f=\sum^{\bar m}_{m=0} \sqrt{p^{\rm th}_{\rm
T}(m)\, p_{\rm T}(m)}$, in which $p^{\rm th}_{\rm T }(m)$ and $p_{\rm T}(m)$ are the
theoretical and experimental distributions, respectively, and the sum is extended up to
the maximum detected photon number, $\bar m$, above which both $p^{\rm th}_{\rm T }(m)$
and $p_{\rm T}(m)$ become negligible. For all data displayed in
Fig.~\ref{Fig:statistics27} the fidelity is rather satisfactory.

\begin{figure}[tb]
\includegraphics[width=0.5\columnwidth]{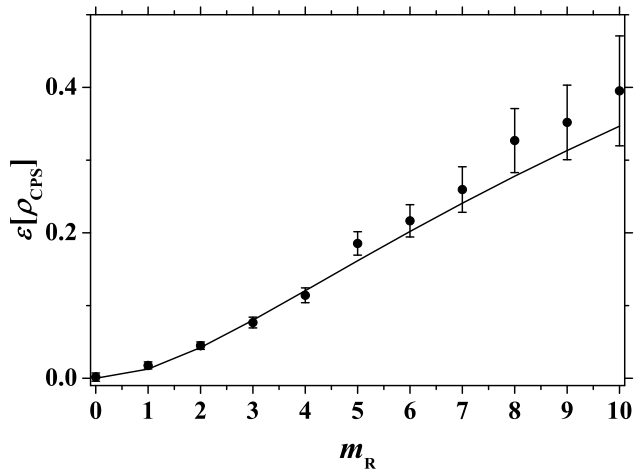}
\caption{\label{Fig:nonG27} Lower bound $\varepsilon[\varrho_{\rm CPS}]$
for the nonG measure $\delta[\varrho_{\rm CPS}]$ as a function of the conditioning value $\mR$
for $M_{\rm T}=1.254$. }
\end{figure}

Finally, the behavior of the lower bound for the nonG measure as a function of the
conditioning value $\mR$ (Fig.\ref{Fig:nonG27}) predicted by the theory (line) is well
reproduced by the experimental data (dots).
In particular, it is worth noting that the value of $\varepsilon[\varrho_{\rm CPS}]$
increases at increasing the conditioning value.

\begin{figure}[tb]
\includegraphics[width=0.5\columnwidth]{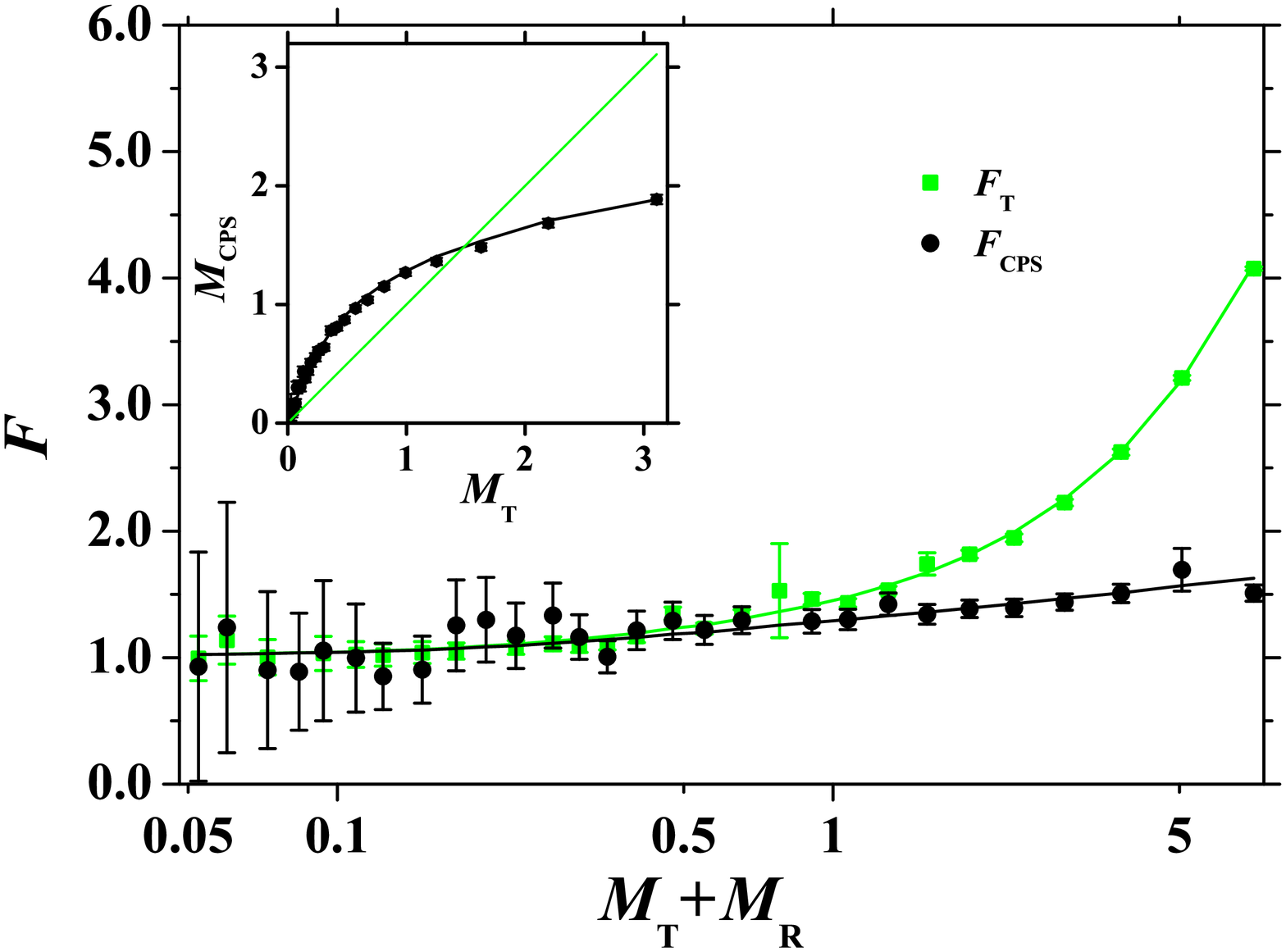}
\caption{\label{Fig:FanoCPS} (Color online)  Log-Linear plot of the Fano
factors $\FCPS$ of the CPS states (black dots) and of the
unconditional states $\FT$ (green squares) as functions of the total mean
detected photons $M_{\rm T} + M_{\rm R}$ for conditioning $\mR=2$.
The solid lines refer to the corresponding  theoretical curves.
The inset shows the mean number of detected photons $M_{\rm CPS}$ of
the CPS states as a function of the mean number of detected photons
$M_{\rm T}$ of the unconditional states: experimental data (black dots)
and theoretical curve (solid line). The green line refers to the mean photon
number $M_{\rm T}$ of the unconditional states.}
\end{figure}

As an example of the second approach, we consider the CPS states obtained by choosing
$\mR = 2$ as the conditioning value for different values of $\Nth$. In the inset of
Fig.~\ref{Fig:FanoCPS} the mean number of photons of the CPS states is plotted together
with the mean number of photons of the initial states measured in the transmitted arm: it
is interesting to notice that the values of $M_{\rm CPS}$ actually approach the
conditioning value $\mR=2$. Again, the experimental results (dots) are well superimposed
to the theoretical curves, calculated starting from Eq.~(\ref{rho:CPS}) with the measured
mean values.

Figure~\ref{Fig:FanoCPS} also shows the comparison between the Fano factor of the
unconditional states $\FT$ (green squares) and that of the CPS  states $\FCPS$ (black
dots): as expected from theory the conditional states preserve the super-Poissonian
nature of the incoming states, though with a smaller value of the Fano factor ($\FT \ge
\FCPS \ge 1$).

\begin{figure}[tb]
\includegraphics[width=0.5\columnwidth]{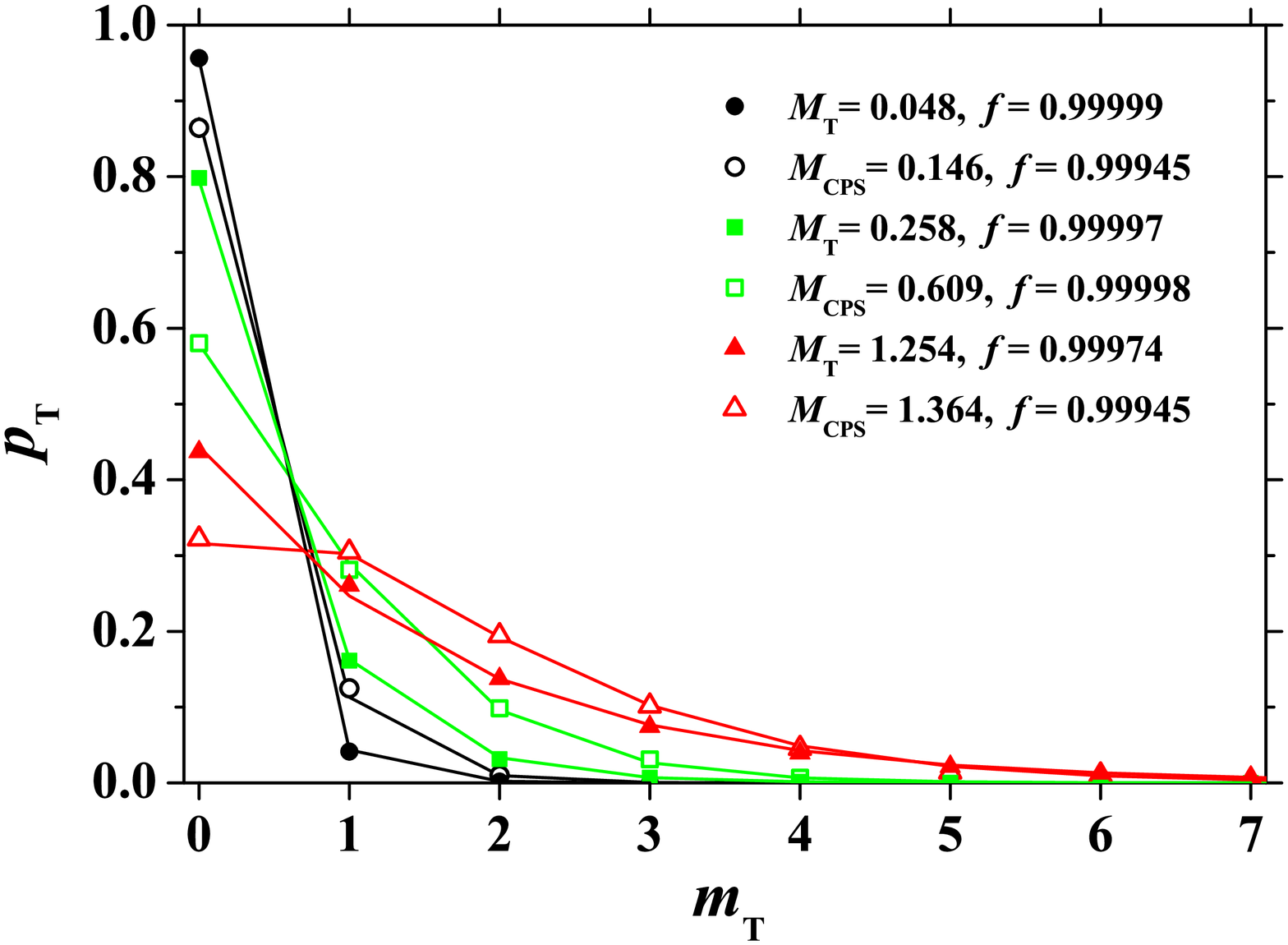}
\caption{\label{Fig:statisticsCPS} (Color online) Reconstructed
photon-number distributions for three different examples (black,
green, red) of unconditional states (full symbols) and for the
corresponding CPS states (empty symbols) with $\mR=2$.  The theoretical curves are
plotted as lines according to the same choice of colors. The corresponding
fidelity $f$ is also reported.}
\end{figure}
In Fig.~\ref{Fig:statisticsCPS} we show three examples of conditional-state photon
distributions for different values of the total incident intensity. For each example, we
plot both the original thermal distribution (full symbols) and that of the conditional
state (empty symbols). The agreement with the corresponding theoretical predictions
(lines) is again testified by the high value of the fidelities.

\begin{figure}[tb]
\includegraphics[width=0.5\columnwidth]{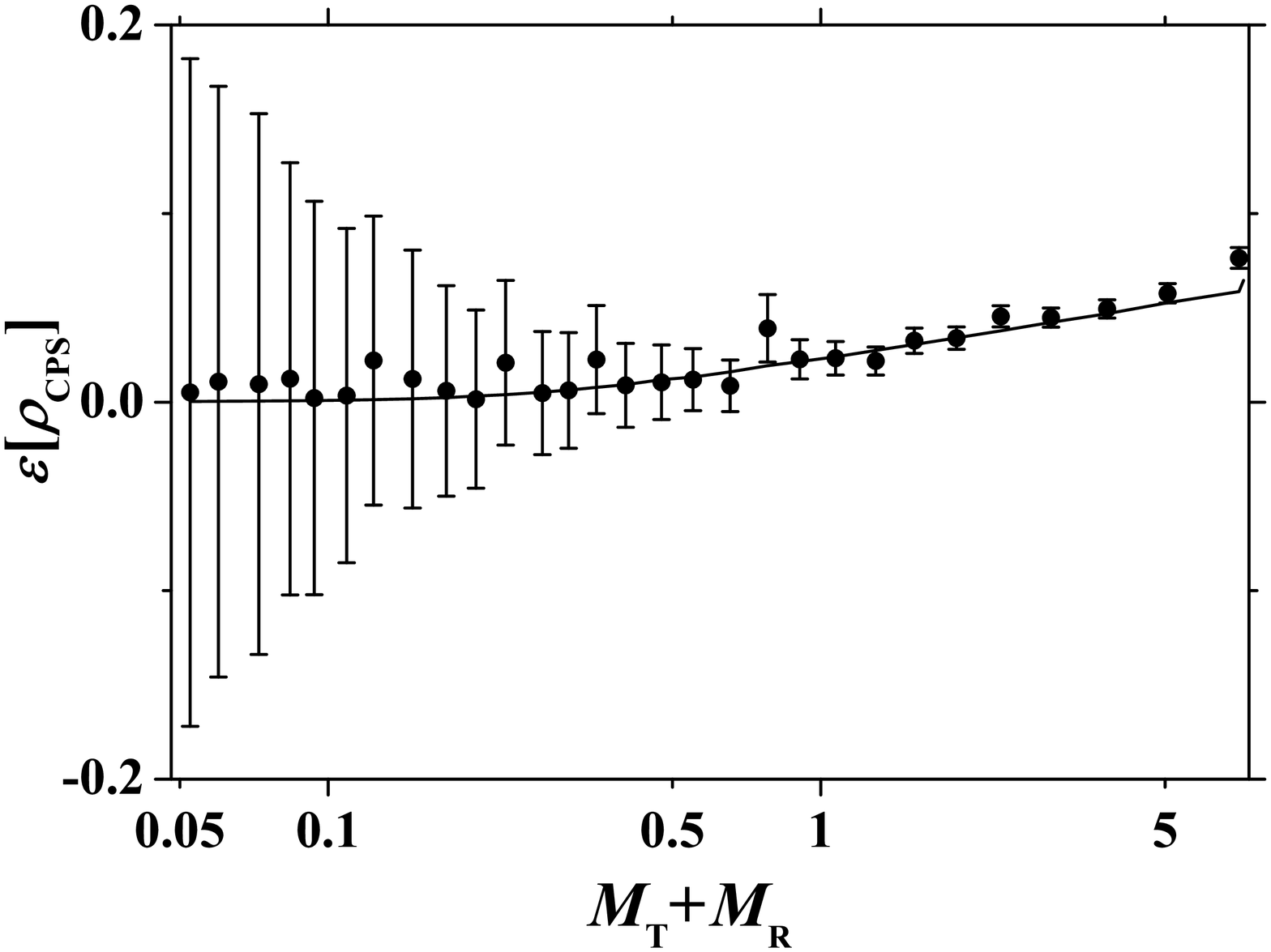}
\caption{\label{Fig:nonGCPS} Log-linear plot
of the lower bound $\varepsilon[\varrho_{\rm CPS}]$ for the non-Gaussianity $\delta[\varrho_{\rm IPS}]$ as a function of the total mean
detected photons $M_{\rm T}+M_{\rm R}$ and for $\mR=2$:
experimental data (dots) and theoretical curve (solid line).}
\end{figure}
We plot in Fig.~\ref{Fig:nonGCPS} the lower bound for the non-Gaussianity
$\varepsilon[\varrho_{\rm CPS}]$ as a function of the total mean detected photons (see
Fig.~\ref{Fig:nonGCPS}) together with the expected theoretical results.


\subsection{IPS non-Gaussian states}\label{s:IPS:exp}

Here we consider the scenario in which an on/off Geiger-like detector measures the
reflected part of the input signal. In particular, as described in
Section~\ref{s:IPS:th}, we are interested in studying the properties of the state
produced in the transmitted arm of the PBS whenever the detector placed in the reflected
arm clicks.  To this aim, we performed a set of measurements by fixing the transmissivity
of the PBS $\tau=0.5$ and changing the mean intensity of the light impinging on the PBS.

\begin{figure}[tb]
\includegraphics[width=0.5\columnwidth]{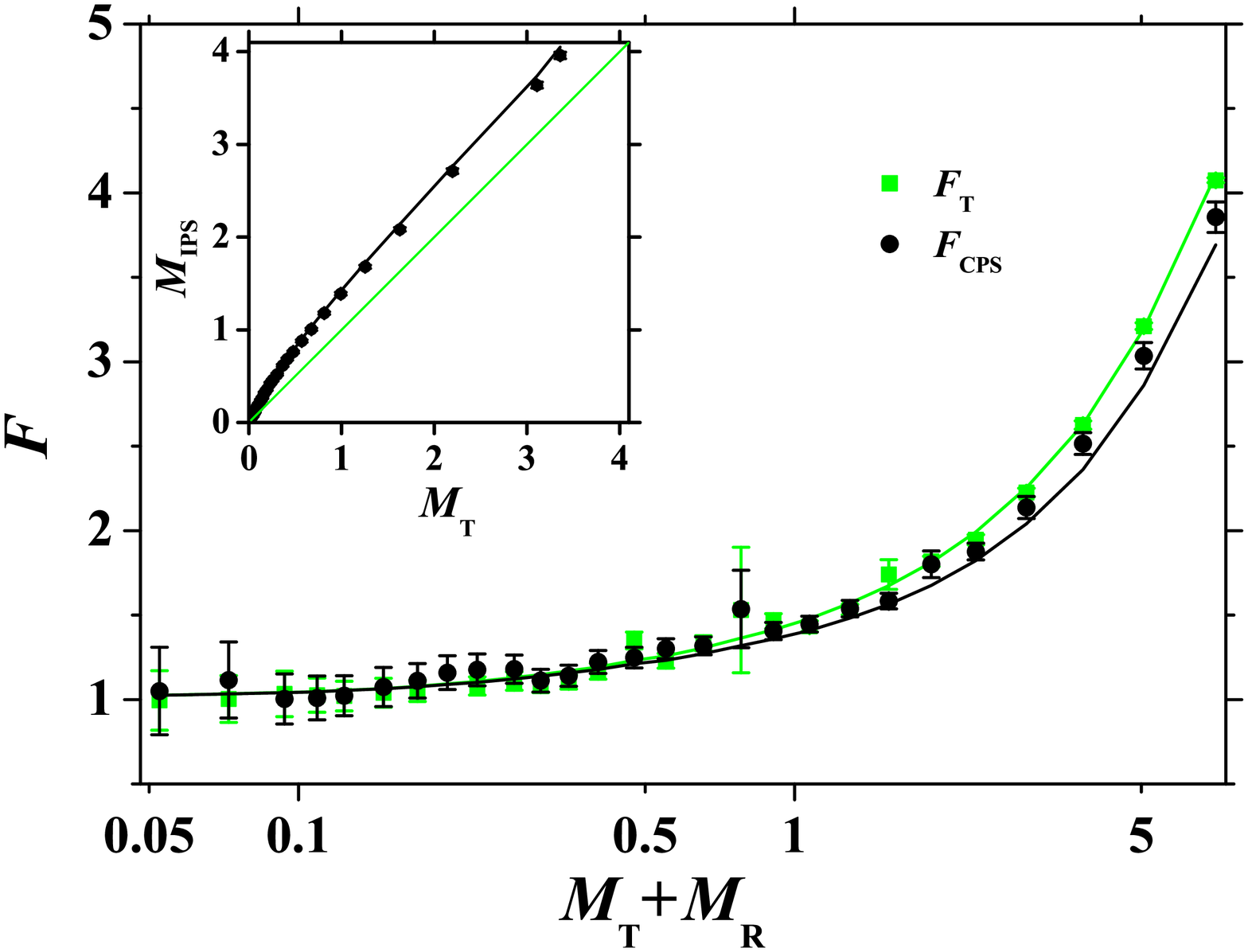}
\caption{\label{Fig:FanoIPS} (Color online) Log-linear plot of the Fano
factors $\FIPS$ of the IPS states (black dots) and of the unconditional states
(green squares) as functions of the total mean
detected photons
$M_{\rm T}+M_{\rm R}$. The solid lines refer to the corresponding theoretical curves.
The inset shows the mean number of detected
photons $M_{\rm IPS}$ of the IPS states as a function of the mean
number of detected photons $M_{\rm T}$ of the unconditional states:
experimental data (black dots) and theoretical curve (solid line).
The green line refers to the mean photon number $M_{\rm T}$ of the unconditional
states.}
\end{figure}

In the inset of Fig.~\ref{Fig:FanoIPS} we plot the mean number of detected photons
$M_{\rm IPS}$ of the IPS states as a function of the mean number of detected photons
$M_{\rm T}$ of the unconditional thermal states (black dots) together with the
theoretical prediction (solid line) according to Eq.~(\ref{N:IPS}). We note that the
effect of the conditioning operation is to increase the mean value of the original state.
As described in Section~\ref{s:IPS:th}, another quantity to characterize the IPS state is
the Fano factor $\FIPS$: to better appreciate the difference between the unconditional
states and the corresponding conditional ones, we plot in the same figure (see
Fig.~\ref{Fig:FanoIPS}) the corresponding Fano factors as functions of the total mean
detected photons (symbols). For each set of experimental results we also plot the
theoretical behaviors (lines): analogously to the conditional case, we have $\FT \ge
\FIPS \ge 1$.

\begin{figure}[tb]
\includegraphics[width=0.5\columnwidth]{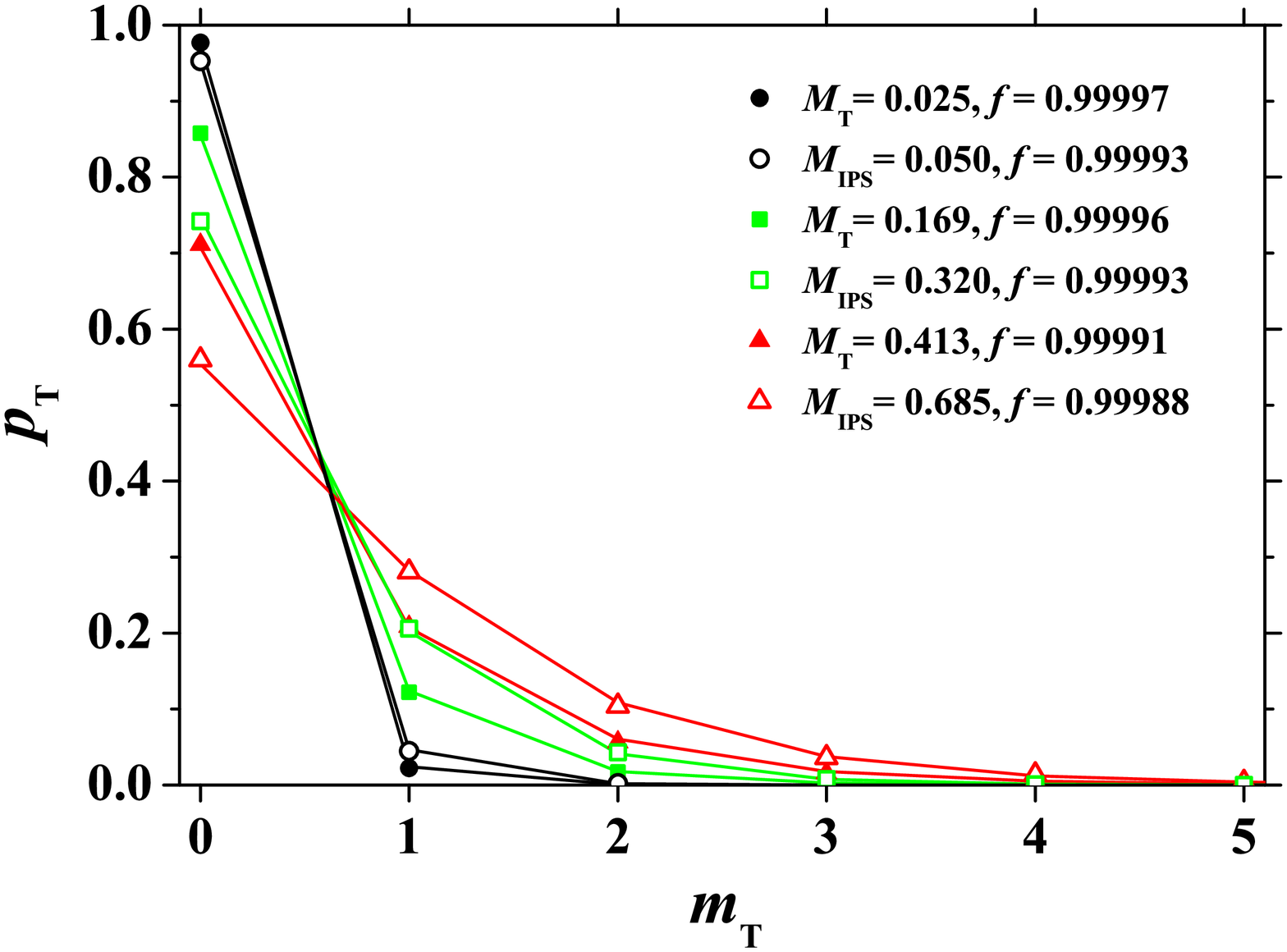}
\caption{\label{Fig:statisticsIPS} (Color online) Reconstructed
photon-number distributions for three different examples (black, green, red)
of unconditional states (full symbols) and for the corresponding
IPS states (empty symbols). The theoretical curves are
plotted as lines according to the same choice of colors. The
fidelity $f$ is also reported.}
\end{figure}
%
In Fig.~\ref{Fig:statisticsIPS} we show the reconstruction of the \emph{detected} photons
distribution $p_{\rm T}(\mT)$ of both the unconditional (full symbols) and the
conditional states (empty symbols) for three different mean values (black, green, red) of
the incident intensity. The agreement with the corresponding theoretical distributions
(colored lines), calculated with the measured mean values, can be checked by evaluating
the fidelity, as reported in Fig.~\ref{Fig:statisticsIPS}.

\begin{figure}[tb]
\includegraphics[width=0.5\columnwidth]{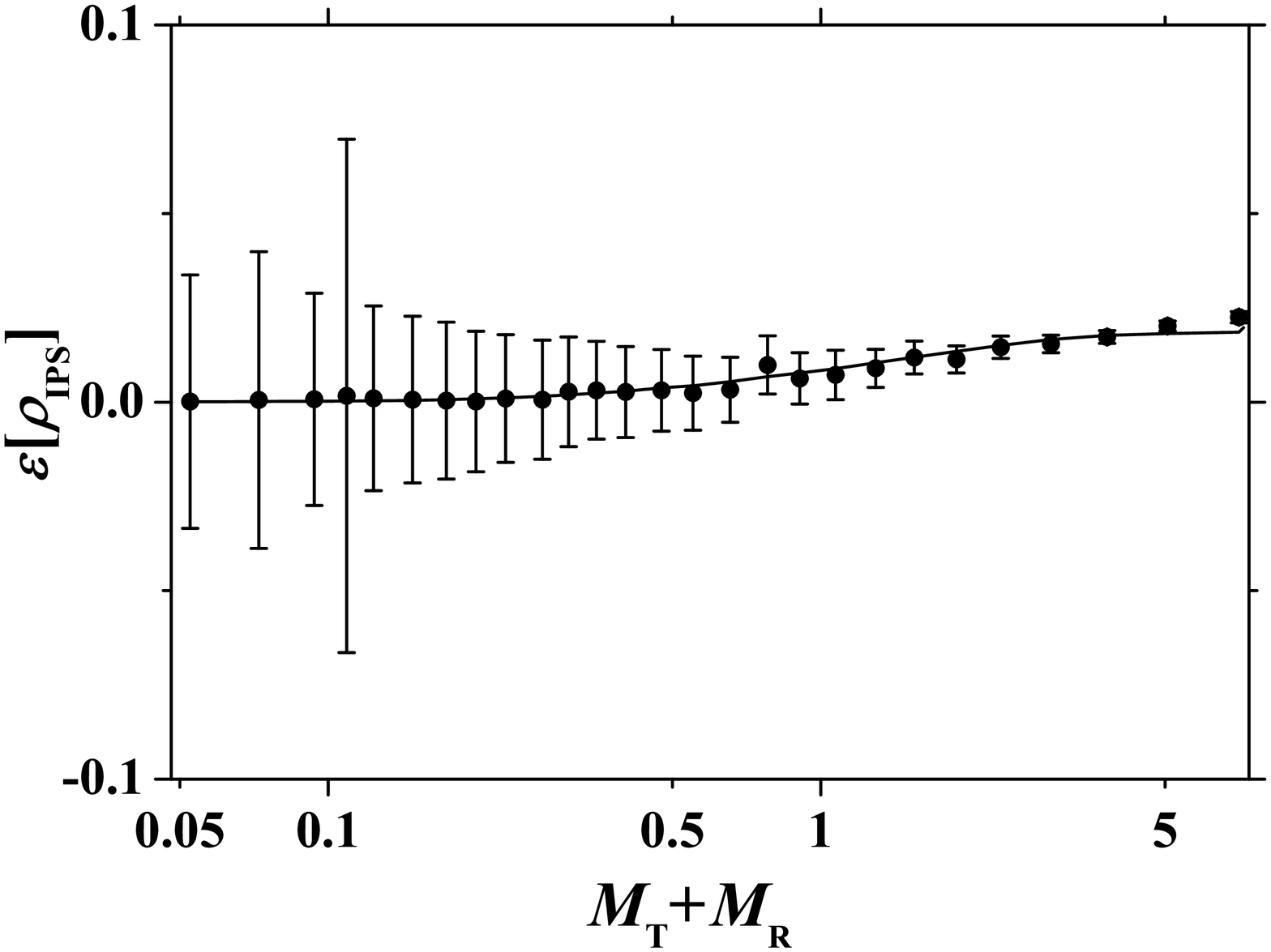}
\caption{\label{Fig:nonGIPS} Log-linear
plot of the lower bound $\varepsilon[\varrho_{\rm IPS}]$ for the non-Gaussianity $\delta[\varrho_{\rm IPS}]$
as a function of the total mean detected photons $M_{\rm T} + M_{\rm R}$:
experimental data (dots) and theoretical curve (line).}
\end{figure}
Finally, in Fig.~\ref{Fig:nonGIPS} we plot the lower bound for the nonG measure
$\varepsilon[\varrho_{\rm IPS}]$ as a function of the total mean detected photons. The
correspondence between the experimental results (dots) and the theoretical prediction
(line) is good.  Note that $\varepsilon[\varrho_{\rm IPS}]$ increases as the mean number
of detected photons increases: this allows the generation of highly populated
non-Gaussian states.

\section{Concluding remarks}\label{s:concl}

In this paper we have discussed in detail, both from a theoretical and an experimental
point of view, a setup based on a single beam splitter and two photon-number resolving
detectors to subtract photons from an incoming state and, thus, to generate non-Gaussian
states starting from Gaussian ones. In order to show the reliability of our setup, we
used (Gaussian) thermal states as inputs and completely characterized the conditional
non-Gaussian outgoing states. In our analysis, we have adopted two possible detection
schemes: the first one is based on the conclusive photon subtraction (CPS), whereas the
second one on the inconclusive photon subtraction. In particular, we have demonstrated,
as one may expect, that the non-Gaussianity of a state increases by increasing either the
intensity of the input states or the conditioning value in the CPS scenario. This last
condition requires photon-counting detectors endowed with a good linear response, such as
those we used in our experiment.

The use of a class of diagonal states in the photon number basis (the thermal ones),
allows us to obtain a high degree of control between the analytical theoretical
expectations and the experiment, which is a relevant point in view of further
investigations. In particular, we are planning to apply our schemes to more exotic
classical states, such as the phase-averaged coherent states \cite{bon:ASL:09}. These are
characterized by a non-Gaussian nature themselves, and, thus, the possibility to perform
conditional, non-Gaussian measurements on them becomes particularly intriguing. Moreover,
in this case analytical calculations may be carried out only to a certain extent: this is
a clear example in which the reliability of the setup is a key point, as theoretical
expectations are limited to numerical results.

Though we only focused on classical states, our experimental procedure opens the way to
further developments toward the generation and engineering of more sophisticated quantum
states by considering non-classical states as the inputs \cite{geno:PS:10,twbrd}, which
may be useful for quantum information protocols involving non-Gaussian states, such as
entanglement distillation protocols \cite{browne:PRA:03,eisert:AOP:04,taka:09}.

\acknowledgments
We would like to thank M.~G.~A.~Paris for his encouragement, advises
and careful, critical reading of this manuscript.
SO acknowledges useful discussions with V.~C.~Usenko and S.~Cialdi.
This work has been partially supported by the CNR-CNISM agreement.

\appendix
\section{Experimental lower bound for the non-Gaussianity} \label{appA}
For a single-mode state diagonal in the Fock basis, i.e., $\varrho=\sum_n p_n
|n\rangle\langle n|$, the nonG measure \cite{non:g} is given by
\begin{align}
\delta[\varrho] &= S[\nu(N)] - S(\varrho) = S[\nu(N)] + \sum_n p_n \log p_n, \label{eq:nonG}
\end{align}
where $\nu(N)$ is a thermal state with mean photon number $N=\sum_n n \, p_n$. Being
based on the knowledge of the actual photon distribution $p_n$, the calculation of
$\delta[\varrho]$ requires measuring with an ideal (i.e., with unit quantum efficiency)
photon-number resolving detector. In the presence of inefficient detection, one can only
retrieve the {\em detected} photon number distribution $q_m = \Tr [ \varrho \Pi_m(\eta)
]$, where $\Pi_m(\eta)$ is given in Eq.~(\ref{POVM:real}) and $\eta$ is the quantum
efficiency. Nevertheless, in the following we will show that the quantity
\begin{align}
\varepsilon[\varrho] = S[\nu(M)] + \sum_m q_m \log q_m, \label{eq:nonGP}
\end{align}
where $M=\sum_m m \, q_m=\eta N$, is a lower bound for the real non-Gaussianity
$\delta[\varrho]$, i.e., $\varepsilon[\varrho] \le \delta[\varrho]$. Note that since
$\varepsilon[\varrho]$ depends only on $q_m$, it can be calculated starting from the
experimental results.

The inefficient photodetection process can be described by mixing the quantum state
$\varrho$ with the vacuum at a BS having transmissivity $\eta$ followed by perfect
detection on the transmitted beam, thus obtaining
\begin{align}
q_m = \Tr [ \mathcal{E}(\varrho) |m\rangle\langle m|],
\end{align}
where $\mathcal{E}(\varrho)=\Tr_2[ U_{BS}(\eta) \varrho \otimes |0\rangle\langle 0|
U_{BS}^{\dagger}(\eta)] $ is the lossy Gaussian channel. Since $\varrho$ is diagonal in
the Fock basis, $\mathcal{E}(\varrho)$ is still diagonal
\begin{align}
\mathcal{E}(\varrho) = \sum_n p_n \mathcal{E}(|n\rangle\langle n|)
= \sum_m q_m |m\rangle\langle m|,
\end{align}
in which $q_m = \sum_{n=m}^{\infty} p_n B_{n,m}(\eta)$. To obtain $\mathcal{E}(\varrho)$
we used $\mathcal{E}(|n\rangle\langle n|) = \sum_{l=0}^n  B_{n,l}(\eta) |l\rangle \langle
l|$, with $ B_{n,l}(\eta)$ defined in Eq.~(\ref{POVM:real}). By using
Eq.~(\ref{eq:nonG}), we obtain
\begin{align}
\delta[{\cal E}(\varrho)] &= S[\nu(M)] + \sum_n q_n \log q_n = \varepsilon[\varrho].
\end{align}
As the nonG measure $\delta[\varrho]$ is non-increasing under Gaussian maps \cite{non:g},
we finally get
\begin{align}
\varepsilon[\varrho] = \delta[\mathcal{E}(\varrho)] \leq
\delta[\varrho].
\end{align}
Summarizing, given a quantum state $\varrho$, diagonal in the Fock basis, we can measure
the probability distribution of the detected phtotons $q_m$ and evaluate
Eq.~(\ref{eq:nonGP}) as a lower bound for the actual non-Gaussianity $\delta[\varrho]$.




\begin{thebibliography}{50}
\bibitem{DC:note} The most general POVM describing a photon counting device also takes
    into account the presence of the dark counts; however, for the photodetectors we used
    in the experimental demonstration the number of the dark counts is negligible.
\end{thebibliography}
\end{document}